\documentclass[floats,floatfix,showpacs,amssymb,prd,twocolumn,superscriptaddress,nofootinbib]{revtex4-1}

\usepackage{graphicx, epsf, epsfig, amssymb}
\usepackage{bm}
\usepackage{longtable}
\usepackage[usenames,dvipsnames]{xcolor}
\usepackage{color}
\usepackage[breaklinks]{hyperref}
\usepackage{amsfonts,amsmath,amssymb,mathrsfs}

\def\be{\begin{equation}}
\def\ee{\end{equation}}
\def\beq{\begin{eqnarray}}
\def\eeq{\end{eqnarray}}
\def\nn{\nonumber}

\begin{document}

\title{Partially massless gravitons do not destroy general relativity black holes}

\author{Richard Brito}
\email{richard.brito@ist.utl.pt}
\affiliation{CENTRA, Departamento de F\'{\i}sica, Instituto Superior
  T\'ecnico, Universidade T\'ecnica de Lisboa - UTL, Avenida~Rovisco Pais
  1, 1049 Lisboa, Portugal}

\author{Vitor Cardoso}
\affiliation{CENTRA, Departamento de F\'{\i}sica, Instituto Superior
  T\'ecnico, Universidade T\'ecnica de Lisboa - UTL, Avenida~Rovisco Pais
  1, 1049 Lisboa, Portugal}
\affiliation{Faculdade de F\'{\i}sica, Universidade Federal do Par\'a, 66075-110, Bel\'em, Par\'a, Brazil}
\affiliation{Perimeter Institute for Theoretical Physics
Waterloo, Ontario N2J 2W9, Canada
}
\affiliation{Department of Physics and Astronomy, The University of Mississippi, University, MS 38677, USA.
}

\author{Paolo Pani}
\affiliation{CENTRA, Departamento de F\'{\i}sica, Instituto Superior
  T\'ecnico, Universidade T\'ecnica de Lisboa - UTL, Avenida~Rovisco Pais
  1, 1049 Lisboa, Portugal}
\affiliation{Institute for Theory $\&$ Computation, Harvard-Smithsonian
CfA, 60 Garden Street, Cambridge, MA, USA}
%


\begin{abstract}
Recent nonlinear completions of Fierz-Pauli theory for a massive spin-2 field include nonlinear massive gravity and bimetric theories. The spectrum of black-hole solutions in these theories is rich, and comprises the same vacuum solutions of Einstein's gravity 
enlarged to include a cosmological constant. 
It was recently shown that Schwarzschild (de Sitter) black holes in these theories are generically unstable against spherical perturbations. Here we show that a notable exception is partially massless gravity, where the mass of the graviton is fixed in terms of the cosmological constant by $\mu^2=2\Lambda/3$ and a new gauge invariance emerges. We find that general-relativity black holes are stable in this limit. Remarkably, the spectrum of massive gravitational perturbations is isospectral.
\end{abstract}

\pacs{04.50.Kd, 04.50.-h, 04.70.Bw, 04.25.dg}
\maketitle

\section{Introduction}
Triggered by several circumstances, the last couple of years have witnessed a flurry of activity on theories with a propagating massive graviton.
Such theories come under different flavors, such as ``nonlinear massive gravity,'' a nonlinear generalization of the linear Fierz-Pauli theory~\cite{deRham:2010ik,deRham:2010kj,Hassan:2011hr}, and ``bimetric theories of gravity'' which propagate two dynamical spin-2 fields~\cite{Isham:1971gm,Salam:1976as,Hassan:2011zd}.
With a very special status, ``partially massless (PM) theories'' have also been considered for which the graviton mass is constrained to take a specific value dictated by the cosmological constant and a new gauge symmetry emerges~\cite{Deser:1983mm,Deser:2001pe,Hassan:2012gz,Deser:2012qg,Deser:2013uy,deRham:2012kf,deRham:2013wv}.

The spectrum of vacuum solutions in these theories is rich, and it comprises the same black-hole (BH) solutions of Einstein's gravity~\cite{Hassan:2011zd,Volkov:2013roa}. However, it was very recently shown that general-relativity BHs are dynamically unstable in these theories~\cite{Babichev:2013una,Brito:2013wya}.
The instability is due to a propagating spherically symmetric degree of freedom and affects BHs with or without a cosmological constant~\cite{Brito:2013wya}. 
As discussed below, such degree of freedom is absent in PM gravity~\cite{Higuchi1987397,Deser:1983mm,Deser:2001pe}, so one might wonder whether Schwarzschild de Sitter (SdS) BHs are stable in such theories.
The purpose of this study is to show that this is the case.

\section{Setup}
Let us consider the propagation of a massive spin-2 field on a curved spacetime. We assume the  background to be a vacuum solution of Einstein's equations with a cosmological constant $\Lambda$ [hereafter we use $G=c=\hbar=1$ units]:
\be
\bar{R}_{\mu\nu}-\frac{1}{2}\bar{R}\bar{g}_{\mu\nu}+\Lambda \bar{g}_{\mu\nu}=0\,,\label{background_eqs}
\ee
so that $\bar{R}=4\Lambda$, $\bar{R}_{\mu\nu}=\Lambda \bar{g}_{\mu\nu}$ and bar quantities refer to the background. The most general linearized action describing a massive spin-2 field minimally coupled to gravity in this background reads [see Ref.~\cite{Hinterbichler:2011tt} for a review] 
\beq\label{FPaction}
S&=&\frac{1}{16\pi}\int\,d^4x\,\sqrt{|\bar{g}|}\left[-\frac{1}{2}\nabla_{\alpha}h_{\mu\nu}\nabla^{\alpha}h^{\mu\nu}+\nabla_{\alpha}h_{\mu\nu}\nabla^{\nu}h^{\mu\alpha}\right.\nn\\
&&-\nabla_{\mu}h\nabla_{\nu}h^{\mu\nu}+\frac{1}{2}\nabla_{\mu}h\nabla^{\mu}h+\frac{\bar{R}}{4}\left(h_{\mu\nu}h^{\mu\nu}-\frac{1}{2}h^2\right)\nn\\
&&\left.-\frac{\mu^2}{2}\left(h_{\mu\nu}h^{\mu\nu}- h^2\right)\right]\,,
\eeq
where $h=\bar{g}_{\mu\nu}h^{\mu\nu}$. From this action the field equations read:
\be \bar{\mathcal{E}}_{\mu\nu}^{\rho\sigma}h_{\rho\sigma}-\Lambda\left(h_{\mu\nu}-\frac{h}{2}\bar{g}_{\mu\nu}\right)+\frac{\mu^2}{2}\left(h_{\mu\nu}-\bar{g}_{\mu\nu}h\right)=0\,,\label{eq1}
\ee
where we have defined the operator
\beq -2\bar{\mathcal{E}}_{\mu\nu}^{\rho\sigma}&=&(\delta_\mu^\rho\delta_\nu^\sigma-\bar{g}_{\mu\nu}\bar{g}^{\rho\sigma})\bar{\square}+\bar{g}^{\rho\sigma}\bar{\nabla}_\mu\bar{\nabla}_\nu-2\delta_{(\mu}^\rho\bar{\nabla}^\sigma\bar{\nabla}_{\nu)}\nn\\
 &&+\bar{g}_{\mu\nu}\bar{\nabla}^\rho\bar{\nabla}^\sigma\,,\label{operator}
\eeq
and index round brackets denote symmetrization.

Finally, by taking the divergence and the trace of Eq.~\eqref{eq1}, and using the tensorial relation
\be
(\bar \nabla_c\bar \nabla_d - \bar \nabla_d\bar \nabla_c)h_{ab} = \bar R_{aecd}h^e\,_b + \bar R_{becd}h_a\,^e\,,  \label{commutator}
\ee
the linearized equations reduce to the system:
\be
\label{eqmotioncurved}
\left\{
\begin{array}{l}
 \bar\Box h_{\mu\nu}+2 \bar R_{\alpha\mu\beta\nu} h^{\alpha\beta}-\mu^2 h_{\mu\nu}=0\,,\\
 \mu^2\bar\nabla^{\mu}h_{\mu\nu}=0\,,\\
 \left(\mu^2-{2\Lambda}/{3}\right)h=0\,.
\end{array}\right.
\ee
Since we are interested in astrophysical scenarios, hereafter we assume $\Lambda\geq0$.
The linearized theory generically propagates five degrees of freedom, corresponding to the healthy helicities of a massive spin-2 field, and it does not contain Boulware-Deser ghosts~\cite{Boulware:1973my}.
The value 
\be
\mu^2=2\Lambda/3\label{PM_limit}
\ee
plays a special role in asymptotically dS spacetimes and it is known as Higuchi limit~\cite{Higuchi1987397,Deser:1983mm,Deser:2001pe}. When $\mu^2<2\Lambda/3$, the helicity-0 mode becomes itself a ghost, whereas when $\mu^2>2\Lambda/3$ all propagating degrees of freedom are physical. In the Higuchi limit the tracelessness of $h_{\mu\nu}$ is not enforced by the field equations~\eqref{eqmotioncurved} and an extra gauge symmetry can be used to eliminate the helicity-0 mode. In this particular case, known as PM gravity~\cite{Deser:2012qg,Deser:2013uy}, the graviton propagates only four helicities.  The attempt to find a nonlinear completion of PM gravity using the framework of nonlinear massive gravity has been shown to suffer from some obstructions~\cite{deRham:2013wv,Deser:2013uy}, nevertheless advances in finding a consistent full nonlinear theory of PM gravity have been recently made in the context of bimetric theories~\cite{Hassan:2012gz,Hassan:2012rq,Hassan:2013pca,Akrami:2013km}.

Although the setup above describes a ``probe'' spin-2 field minimally coupled to standard gravity, it is in fact very generic. 
Several nonlinear completions of massive gravity --~including the recent proposals of Refs.~\cite{deRham:2010ik,deRham:2010kj,Hassan:2011hr}~-- admit general-relativity background solutions as in Eq.~\eqref{background_eqs}. Furthermore, the linearized field equations can be written as Eq.~\eqref{eq1}, where $\mu$ and $\Lambda$ depend on the specific parameters of the theory~\cite{Hassan:2011zd,Brito:2013wya}. Finally, Eqs.~\eqref{eqmotioncurved} are the only field equations that consistently describe a massive spin-2 field coupled to gravity in generic backgrounds with constant curvature~\cite{Buchbinder:1999ar}.

\subsection{The Schwarzschild-de Sitter geometry}
The most general static solution of Eq.~\eqref{background_eqs} is the SdS spacetime, described by
$d\bar{s}^2 = -f\, dt^2 + f^{-1}\, dr^2 + r^2 d\Omega^2$,
where~\cite{Cardoso:2003sw}
\begin{equation}
f=\frac{\Lambda}{3 r}\, (r-r_b)(r_c-r)(r-r_0)\,,
\label{eq:f_def}
\end{equation}
with $r_0 = -(r_b + r_c)$, $r_b$ and $r_c>r_b$ being the BH horizon and the cosmological horizon, respectively.
The cosmological constant can be expressed as $3/\Lambda={r_b}^2 + r_b r_c + {r_c}^2$
and the spacetime has mass $M= \Lambda r_b r_c (r_b + r_c)/6$.
The surface gravity $\kappa_b$ associated with the BH horizon reads~\cite{Cardoso:2003sw}
\begin{equation}
\kappa_b\equiv\frac{f'(r_b)}{2} = \frac{ \Lambda(r_c-r_b)(r_b-r_0) }{ 6 r_b }.
\label{surface}
\end{equation}
%

\subsection{Field equations in PM gravity}
%
\begin{figure*}[htb]
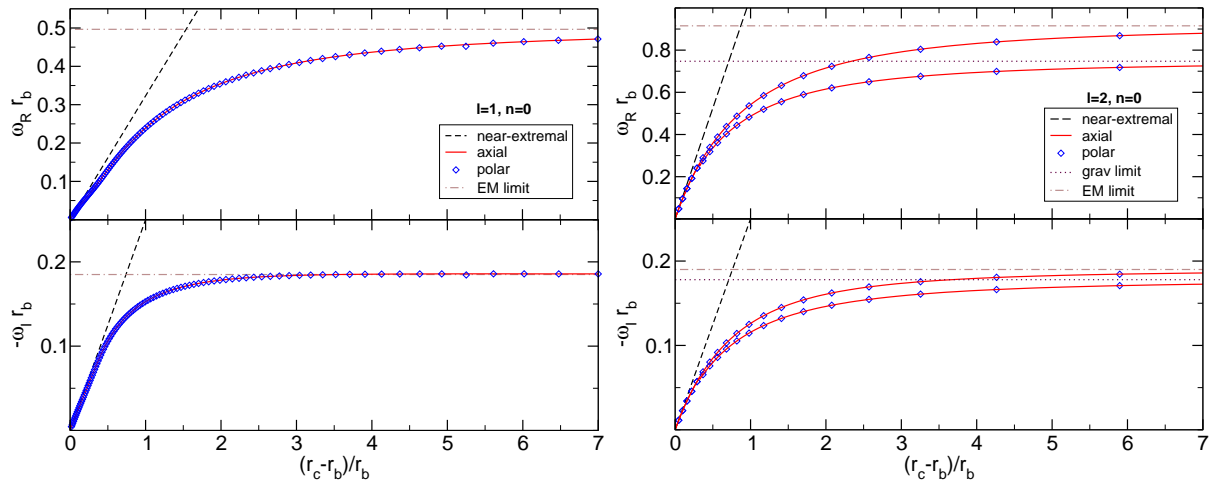

\begin{center}
\begin{tabular}{cc}
\epsfig{file=isospectrality_l1_PM_Re_Im_2.eps,width=7.9cm,angle=0,clip=true}&
\epsfig{file=isospectrality_l2_PM_Re_Im_2.eps,width=7.9cm,angle=0,clip=true}
\end{tabular}
\caption{Left panel: Real (top panel) and imaginary (bottom panel) part of the fundamental dipole mode of a SdS BH in PM gravity. The fundamental mode is the same for both the axial and polar sectors to within numerical accuracy. Similar results hold for the overtones. Numerical results are compared to the analytical expression~\eqref{finalsclarelectr}. The leftmost and rightmost parts of the $x-$axis are the extremal and the general-relativity, asymptotically flat limit ($\Lambda\propto\mu^2\to0$), respectively. In the $\Lambda\propto\mu^2\to0$ limit the $l=1$ modes approach the electromagnetic quasinormal modes of a Schwarzschild BH~\cite{Berti:2009kk}. Right panel: same for $l=2$ modes. For $l>1$ there are two families of modes which, in the $\Lambda\propto\mu^2\to0$ limit approach the gravitational and the electromagnetic quasinormal modes of a Schwarzschild BH, respectively. 
\label{fig:iso_l1}}
\end{center}
\end{figure*}
In a spherically symmetric background, Eqs.~\eqref{eqmotioncurved} can be conveniently decomposed in a complete basis of tensor spherical harmonics.
Perturbations separate into two sets, which are called ``axial'' and ``polar'' according to their parity. In Fourier space this can be written as~\cite{Berti:2009kk,Brito:2013wya}
\begin{align}
\label{decom}
h_{\mu\nu}(t,r,\theta,\phi)&=\sum_{l,m}\int_{-\infty}^{+\infty}e^{-i\omega t}\left[h^{{\rm axial},lm}_{\mu\nu}(\omega,r,\theta,\phi)\right.\nn\\
&\left.+h^{{\rm polar},lm}_{\mu\nu}(\omega,r,\theta,\phi)\right]d\omega\,,
\end{align}
where
\begin{widetext}
\begin{equation}\label{oddpart}
h^{{\rm axial},lm}_{\mu\nu}(\omega,r,\theta,\phi) =
 \begin{pmatrix}
  0 & 0 & h^{lm}_0(\omega,r)\csc\theta\partial_{\phi}Y_{lm}(\theta,\phi) & -h^{lm}_0(\omega,r)\sin\theta\partial_{\theta}Y_{lm}(\theta,\phi) \\
  * & 0 & h^{lm}_1(\omega,r)\csc\theta\partial_{\phi}Y_{lm}(\theta,\phi) & -h^{lm}_1(\omega,r)\sin\theta\partial_{\theta}Y_{lm}(\theta,\phi) \\
  *  & *  & -h^{lm}_2(\omega,r)\frac{X_{lm}(\theta,\phi)}{\sin\theta} & h^{lm}_2(\omega,r)\sin\theta W_{lm}(\theta,\phi)  \\
  * & * & * & h^{lm}_2(\omega,r)\sin\theta X_{lm}(\theta,\phi)
 \end{pmatrix}\,,
\end{equation}
%
%
\begin{align}\label{evenpart}
h^{{\rm polar},lm}_{\mu\nu}(\omega,r,\theta,\phi)=
\begin{pmatrix}
f(r)H_0^{lm}(\omega,r)Y_{lm} & H_1^{lm}(\omega,r)Y_{lm} & \eta^{lm}_0(\omega,r)\partial_{\theta}Y_{lm}& \eta^{lm}_0(\omega,r)\partial_{\phi}Y_{lm}\\
  * & f(r)^{-1} H_2^{lm}(\omega,r)Y_{lm} & \eta^{lm}_1(\omega,r)\partial_{\theta}Y_{lm}& \eta^{lm}_1(\omega,r)\partial_{\phi}Y_{lm}\\
  *  & *  & \begin{array}{c}r^2\left[K^{lm}(\omega,r)Y_{lm}\right.\\\left. +G^{lm}(\omega,r)W_{lm}\right]\end{array} & r^2  G^{lm}(\omega,r)X_{lm}  \\
  * & * & * & \begin{array}{c}r^2\sin^2\theta\left[K^{lm}(\omega,r)Y_{lm}\right.\\\left.-G^{lm}(\omega,r)W_{lm}\right]\end{array}
\end{pmatrix}\,,
\end{align}
\end{widetext}
where asterisks represent symmetric components, $Y_{lm}\equiv Y_{lm}(\theta,\phi)$ are the scalar spherical harmonics and
\beq
X_{lm}(\theta,\phi)=2\partial_{\phi}\left[\partial_{\theta}Y_{lm}-\cot\theta Y_{lm}\right]\,,
\eeq
\beq
W_{lm}(\theta,\phi)=\partial^2_{\theta}Y_{lm}-\cot\theta\partial_{\theta}Y_{lm}-\csc^2\theta\partial^2_{\phi}Y_{lm}\,.
\eeq

As we showed in a recent work, SdS geometries are {\it generically} unstable against a 
monopole fluctuation~\cite{Brito:2013wya}. In the asymptotically flat case, the instability is equivalent to the Gregory-Laflamme instability~\cite{Gregory:1993vy} of a black string~\cite{Babichev:2013una}. Thus, general-relativity BHs cannot describe static solutions in theories whose linearized equations reduce to Eqs.~\eqref{eqmotioncurved}. 

However, a notable exception to this outcome is given by PM theories, which are defined by the tuning~\eqref{PM_limit}.
In this case, the helicity-0 mode which is responsible for the instability can be gauged away and the theory propagates four degrees of freedom~\cite{Higuchi1987397,Deser:1983mm,Deser:2001pe}.

The approach we developed in Ref.~\cite{Brito:2013wya} can be easily extended to obtain a set of coupled master equations that fully characterize the linear stability properties of the background. The axial sector is described by the following system:
\beq
&&\frac{d^2}{dr_*^2}Q+\left[\omega^2-f\left(\frac{\lambda+4}{r^2}-\frac{16M}{r^3}\right)\right]Q=S_Q\,,\label{oddf1}\\
&&\frac{d^2}{dr_*^2}Z+\left[\omega^2-f\left(\frac{\lambda-2}{r^2}+\frac{2M}{r^3}\right)\right]Z=S_Z\,,\label{oddf2}
\eeq
where $\lambda=l(l+1)$ and we have defined the tortoise coordinate $r_*$ via $dr/dr_*=f$. The functions $Q(r)\equiv f(r)h_1$ and $Z(r)\equiv h_2/r$ are combinations of the axial perturbations
, whereas the source terms are given by
\begin{equation}
 S_Q=(\lambda-2)\frac{2f(r-3M)}{r^3}Z\,,\quad S_Z= \frac{2}{r^2}f\,Q\,.
\end{equation}

The polar sector can be simplified by using an extra gauge symmetry arising in the Higuchi limit~\eqref{PM_limit}. In this case the field equations~\eqref{eqmotioncurved} are invariant under~\cite{Higuchi1987397}
\begin{equation}
 h_{\mu\nu}\to h_{\mu\nu}+\left(\bar\nabla_\mu\bar\nabla_\nu+\frac{\Lambda}{3}\bar g_{\mu\nu}\right)\xi
\end{equation}
where $\xi$ is a generic scalar gauge function of the spacetime coordinates. The symmetry above can be used to enforce $\eta_0\equiv0$ in the decomposition~\eqref{evenpart}.
In this gauge, the polar sector is fully described by a system of two coupled ordinary differential equations:
\beq
\label{polar_eq1}
f^2\frac{d^2 \eta_1}{dr^2}+\hat\alpha_1 \frac{d \eta_1}{dr}+\hat\beta_1 \eta_1 &=& S_{\eta_1}\,,\\
\label{polar_eq2}
f^2\frac{d^2 G}{dr^2}+\hat\alpha_2 \frac{d G}{dr}+\hat\beta_2 G &=& S_G\,,
\eeq
where the source terms are given by
\begin{align}
\label{source1}
S_{\eta_1}&= (\lambda-2)\hat\sigma_1 \frac{d G}{dr}+(\lambda-2)\hat\rho_1 G\,,\\
\label{source3}
 S_G &= \hat\sigma_2 \frac{d \eta_1}{dr}+\hat\rho_2 \eta_1\,.
\end{align}
The coefficients $\hat\alpha_i,\,\hat\beta_i,\,\hat\sigma_i,\,\hat\rho_i$ are radial functions which also depend on $\omega$, $l$, $r_b$ and $r_c$.
%
%

The regular asymptotic solutions of the axial and polar systems are ingoing and outgoing waves, $\Psi\to e^{\mp i\omega r_*}$, at the BH horizon and at the cosmological horizon, respectively [$\Psi$ collectively denotes the master functions $Q$, $Z$, $\eta_1$ and $G$]. The complex eigenfrequencies $\omega=\omega_R+i\omega_I$, that simultaneously satisfy these boundary conditions are called quasinormal modes~\cite{Berti:2009kk}.

\section{Results}
\subsection{The near-extremal SdS geometry}
The above equations are in general not analytically solvable.
Fortunately, the geometry is sufficiently rich that it admits a special limit where
one can indeed considerably simplify the equations and solve them analytically.
This regime is the near extremal SdS BH, defined as the spacetime for which the cosmological horizon $r_c$
is very close (in the $r$ coordinate) to the BH horizon $r_b$, i.e.
$\frac{r_c-r_b}{r_b}\ll 1$. As shown in Ref.~\cite{Cardoso:2003sw}, it is possible
to solve analytically a large class of Schr\"{o}dinger-type equations in this background
by adopting a perturbative approach in powers of $r_c-r_b$.
In the near-extremal limit, one can make the following approximations:~\cite{Cardoso:2003sw}
\be
r_0 \sim -2r_{b}^2\,\,;\,\Lambda\sim r_{b}^{-2};\,\,
M \sim \frac{r_b}{3}\,\,;\,\kappa_b \sim \frac{r_c-r_b}{2r_{b}^2}\,.
\label{approximation1}
\ee
The key point of the approximation is to realize that in the near-extremal regime $r\sim r_b\sim r_c$,
as we are interested only in the region between the two horizons. Then, $r-r_0 \sim r_b -r_0 \sim 3r_0$
and thus
\begin{equation}
 r\sim\frac{r_c e^{2\kappa_b r_*}+r_b}{1+e^{2\kappa_b r_*}}\,,\qquad 
 f\sim\frac{(r_c-r_b)^2}{4r_{b}^2\cosh{(\kappa_b r_*)}^2}\,.
\end{equation}
Finally, the equations for massive gravitational perturbations
of near-extremal SdS geometries reduce to 
\begin{equation}
\frac{d^2 \Psi}{d r_*^{2}} +\left\lbrack\omega^2-\frac{\kappa_{b}^2 U_0}{\cosh{(\kappa_b r_*)}^2}\right\rbrack\Psi=0 \,,
\label{waveequation}
\end{equation}
where $\Psi$ can be any of the metric variables. The potential $U_0$ is the same for the axial metric functions $Q,Z$
but it is different for the polar functions $\eta_1,G$. We find
\begin{equation}\label{U0}
U_0=\left\{ \begin{array}{ll}
            -4/3+\lambda\,,   & {\rm \,axial}\\
            \frac{27 \lambda ^3+36 \lambda  \left(9 r_b^2 \omega ^2-1\right)-16}{3 \left(9 \lambda ^2+12 \lambda +36 r_b^2 \omega ^2+4\right)}\,, & {\rm \,polar\,I}	\\
	\frac{27 \lambda ^3-36 \lambda +72 r_b^2 \omega ^2-16}{3 \left(9 \lambda ^2+12 \lambda +36 r_b^2 \omega ^2+4\right)}\,,   &{\rm \,polar\,II}	
\end{array}\right.
\end{equation}

The potential in (\ref{waveequation}) is the well known
P\"oshl-Teller potential~\cite{Poschl:1933zz}. The solutions of the corresponding Schrodinger-like equations
were studied and they are of the hypergeometric type,
(for details see Section 4.1 in Ref.~\cite{Berti:2009kk}).
The eigenfrequencies are given by~\cite{Cardoso:2003sw,Berti:2009kk} 
\be
\frac{\omega}{\kappa_b}=  -\left(n+\frac{1}{2}\right)i+
\sqrt{U_0-\frac{1}{4}},\quad n=0,1,\dots\label{solution}
\ee
Using Eq.~\eqref{U0}, we obtain
\be
\frac{\omega}{\kappa_b}=  -\left(n+\frac{1}{2}\right)i+\sqrt{l(l+1)-\frac{19}{12}},\quad n=0,1,\dots\label{finalsclarelectr}
\ee
for \emph{both} axial and polar perturbations. Thus, we get the surprising result that in this regime all perturbations are isospectral~\cite{Chandra,Berti:2009kk}.
Because the polar potential $U_0$ is frequency-dependent, we also find a second, spurious, root at $\omega=\pm i\frac{2+3\lambda}{6r_b}$.
It is easy to check analytically that at these frequencies the wavefunction is not regular at one of the horizons, thus it does not belong to the spectrum.

\subsection{Numerical results}
Using two independent techniques (a matrix-valued direct integration and a matrix-valued continued-fraction method~\cite{Pani:2013pma}), we have numerically obtained the quasinormal spectrum for generic SdS geometries, looking explicitly for unstable modes, i.e., modes for which ${\rm Im}(\omega)>0$ and which therefore grow exponentially in time while being spatially bounded.
Our results are summarized in Fig.~\eqref{fig:iso_l1}, where we overplot the near-extremal analytical result (denoted by a black dashed line)
and the $\Lambda\to0$ limit (denoted by horizontal lines). Since $\Lambda\propto\mu^2$, this limit corresponds to the massless limit of PM gravity, i.e. to general relativity.

Except for $l=1$ modes, axial and polar perturbations are grouped in two different families which, in the $\Lambda\propto\mu^2\to0$ limit, reduce to gravitational and electromagnetic modes of a Schwarzschild BH, respectively. 
As predicted by our near-extremal analysis, the two families merge in the $r_c\to r_b$ limit.
For $l=1$ there is only one single family which reduces to the dipole electromagnetic modes of a Schwarzschild BH in the $\Lambda\propto\mu^2\to0$ limit.

An intriguing result, which would merit further study is the fact that axial and polar modes have exactly the \emph{same} quasinormal-mode spectrum
for any value of the cosmological constant, up to numerical accuracy. We were not able to produce an analytical proof of this. Isospectrality guarantees that the entire quasinormal spectrum can be obtained from the axial equations~\eqref{oddf1} and~\eqref{oddf2} only. 

To summarize, our numerical results are in excellent agreement with independent analytical/numerical analysis on two opposite regimes,
the general-relativity limit when the cosmological constant vanishes and the near-extremal limit when the two horizons coalesce.
We found no hints of instabilities in the full parameter space.

\section{Discussion}
In addition to the spectrum of stable modes presented above, we have searched for unstable, exponentially-growing modes and found none.
Thus, our analysis provides solid evidence for the linear stability of SdS BHs in PM gravity. This is in contrast with generic theories of massive spin-2 fields (including massive gravitons) in which  SdS BHs are unstable~\cite{Babichev:2013una,Brito:2013wya}.

From our results and from those of Refs.~\cite{Babichev:2013una,Brito:2013wya} the following interesting picture emerges. 
If a theory of massive gravity allows for the same BH solutions of general relativity, the latter are unstable against spherical perturbations. This is the case for Schwarzschild BHs in any consistent theory of a massive spin-2 field, including the recent nonlinear massive gravity~\cite{deRham:2010ik,deRham:2010kj} and bimetric theories~\cite{Hassan:2011zd}, with or without a cosmological constant. The end-state of the instability is an interesting open problem but it is likely that the instability triggers static BHs to develop ``graviton hair''. 

We have shown here that a notable exception to this picture is represented by PM gravity, which is obtained by enforcing the constraint~\eqref{PM_limit}. The unstable monopole is absent in this theory and a complete analysis of nonspherical modes has revealed no instability. Remarkably, the spectrum of massive gravitational perturbations is isospectral in this theory, which is another piece of evidence for its special role within the family of massive gravities.

\textbf{Note added in proof}: After this work has been accepted, a paper appeared~\cite{Deser:2013gpa} which argues against the existence of a consistent nonlinear PM completion in bimetric theories (see also Ref.~\cite{Fasiello:2013woa}). Modifying the kinetic term of the full non-linear action could be a possible way out as recently proposed in~\cite{Hinterbichler:2013eza,*Folkerts:2011ev}. Note however that our linear analysis applies to \emph{any} nonlinear generalization of the Fierz-Pauli theory.
\begin{acknowledgments}
We thank Stanley Deser for interesting discussions and suggestions.
R.B. acknowledges financial support from the FCT-IDPASC program through the grant SFRH/BD/52047/2012.
V.C. acknowledges
partial financial support provided under the European Union's FP7 ERC Starting
Grant ``The dynamics of black holes: testing the limits of Einstein's theory''
grant agreement no. DyBHo--256667.  Research at Perimeter Institute is supported
by the Government of Canada through Industry Canada and by the Province of
Ontario through the Ministry of Economic Development and Innovation.
P.P acknowledges financial support provided by the European Community through
the Intra-European Marie Curie contract aStronGR-2011-298297.
This work was supported by the NRHEP 295189 FP7-PEOPLE-2011-IRSES Grant, and by
FCT-Portugal through projects CERN/FP/123593/2011.
\end{acknowledgments}
\bibliography{ref}  

\end{document}